\documentclass[aps,prd,reprint,showpacs,superscriptaddress,twocolumn]{revtex4}
\usepackage{amsmath}
\usepackage{amssymb}
\usepackage[breaklinks=true,dvips]{hyperref}
\usepackage{breakurl}
\usepackage{mathrsfs}

\hypersetup{
    colorlinks=true,
    linkcolor=blue,
    citecolor=blue,
    filecolor=blue,
    urlcolor=blue,
    breaklinks=true
  }

\begin{document}

\title{Physical regularization for the spin-1/2 Aharonov-Bohm
problem in conical space}

\author{F. M. Andrade}
\email{fmandrade@uepg.br}
\affiliation{
  Departamento de Matem\'{a}tica e Estat\'{i}stica,
  Universidade Estadual de Ponta Grossa,
  84030-900 Ponta Grossa-PR, Brazil
}

\author{E. O. Silva}
\email{edilbertoo@gmail.com}
\affiliation{
  Departamento de F\'{i}sica,
  Universidade Federal do Maranh\~{a}o, Campus
  Universit\'{a}rio do Bacanga,
  65085-580 S\~{a}o Lu\'{i}s-MA, Brazil
}

\author{M. Pereira}
\email{marciano@uepg.br}
\affiliation{
  Departamento de Matem\'{a}tica e Estat\'{i}stica,
  Universidade Estadual de Ponta Grossa,
  84030-900 Ponta Grossa-PR, Brazil
}

\date{\today}

\begin{abstract}
We examine the bound state and scattering problem of a spin-one-half
particle undergone to an Aharonov-Bohm potential in a conical space in
the nonrelativistic limit. The crucial problem of the
$\delta$-function singularity coming from the Zeeman spin interaction with
the magnetic flux tube is solved through the self-adjoint extension method.
Using two different approaches already known in the literature,
both based on the self-adjoint extension method, we obtain the self-adjoint
extension parameter to the bound state and scattering scenarios in terms
of the physics of the problem. It is shown that such a parameter is the
same for both situations. The method is general and is suitable for
any quantum system with a singular Hamiltonian that has bound and
scattering states.
\end{abstract}


\maketitle

Singularities are very common in quantum mechanics and already have a
long history \cite{PR.1950.80.797}. The first work with $\delta$-like
singularities was in the Kronig-Penny model \cite{PRSLA.1931.130.499}
for the description of the band energy in solid-state physics.
In addition, point interactions
\cite{PRA.2002.66.062712,JPA.2006.39.2493,JOBQSO.2005.7.77} have been of
great interest in various branches of physics for their relevance as
solvable models \cite{Book.2004.Albeverio}. In the Aharonov-Bohm (AB)
effect \cite{PR.1959.115.485} of spin-$1/2$ particles 
\cite{PRD.1989.40.1346,PRL.1990.64.503,IJMPA.1991.6.3119}
a two-dimensional $\delta$ function appears as the mathematical
description of the Zeeman interaction between the spin and the magnetic
flux tube \cite{PRD.1993.48.5935,AP.1996.251.45}. 
Hagen \cite{PRL.1990.64.503} argued that a $\delta$-function
contribution to the potential can not be neglected when the system has
spin, having shown that changes in the amplitude and scattering cross
section arise when the spin of the particle is considered.
Point interactions usually appear in quantum systems in the
presence of topological defects.  A simple but nontrivial example is the
case of a cone rising from an effective geometry 
immersed in several physical systems, such as cosmic strings
\cite{Book.2000.Vilenkin}, defects in elastic media
\cite{AP.1992.216.1}, defects in liquid crystals
\cite{EPJE.2006.20.173}, and so on. In such systems, although the
particle does not have access to the core (defect) region, its wave
function and energy spectrum are truly influenced by it.

Recently, a device was proposed that would detect microstresses in graphene
\cite{Science.2004.306.666} based on a scanning-tunneling-microscopy
setup able to measure AB interferences at the nanometer scale. In this
setup a $\delta$-function scattering potential was considered in the
continuum limit \cite{NaturePhys.2011.7.811}. In
Ref. \cite{PRL.2010.105.206601} it was considered a topological
insulator nanowire  with a magnetic field applied along its length,
focusing on the AB conductance oscillations arising from the surface
states.
The Dirac Hamiltonian of this model takes into account the
spinorial connection that allows us to incorporate topological defects 
(arising from a nontrivial conical geometry) through the metric.
From these studies, such materials could be analyzed through theoretical 
models allowing to include point interactions able to reproduce AB-like 
effects.

In quantum mechanics, singularities and pathological potentials, in
general, are dealt with some kind of regularization procedure. A
common approach to ensure that the wave function in the presence of a
singularity is square-integrable (and therefore might be associated to a
bound state) is to force it to vanish on the singularity. More
appropriately, an analysis based on the self-adjoint extension method
\cite{Book.1975.Reed.II}, broadens the boundary condition possibilities
that still give bound states. The physics of the problem determines
which of these possibilities is the right one, leaving no ambiguities
\cite{PRD.1989.40.1346,NPB.1989.328.140}.
This method has been applied by many authors, in particular, for AB-like
systems \cite{PRD.1989.40.1346,Book.1995.Jackiw,PRD.1994.50.7715,
JMP.1995.36.5453,AP.1996.251.45,AP.1998.263.295}.
However, the results obtained in these works present the most important
results (e.g., energy spectrum, phase shift, S matrix) in terms of an
arbitrary real parameter, the so called self-adjoint extension parameter.

In this article, we describe a general regularization procedure to obtain
the self-adjoint extension parameter, based on the physics of the
spin-1/2 AB system in $(1+2)$-dimensional conical space for both bound
and scattering scenarios.
We take as a starting point the works of Kay-Studer (KS)
\cite{CMP.1991.139.103} and Bulla-Gesztezy (BG) \cite{JMP.1985.26.2520},
both based on the self-adjoint extension method.

The topological defect considered here is a linear quantity that appears
embedded in the metric system $ds^{2}=dr^{2}+\alpha^{2} r^{2}d\varphi^{2}$,
where $r\geq 0$, $0\leq \varphi < 2\pi $, and $\alpha $ is the
parameter which effectively introduces an angular excess or deficit,
identified by $2\pi( 1-\alpha) $. The above metric has a conelike
singularity at $r=0$. In other words, the curvature tensor of this
metric, considered as a distribution, is given by
$R_{12}^{12}=R_{1}^{1}=R_{2}^{2}=2\pi (\alpha -1)\delta(r)/\alpha$,
where $\delta(r) $ is the two-dimensional $\delta$ function
in flat space \cite{SPD.1977.22.312}. This implies a two-dimensional
conical singularity symmetrical in the z axis, which characterizes it as
a linear defect. 

In order to study the dynamics of the particle in a nonflat spacetime,
we should include the spin connection in the differential operator and
define the respective Dirac matrices in this manifold. This system is
governed by the modified Dirac equation in curved space
$[ i\gamma^{\mu }(\partial_{\mu }+\Gamma_{\mu}) -
q\gamma^{\mu}A_{\mu }-M] \psi(x) =0$,
where  $q$ is the charge, $M$ is mass of the particle, $\psi(x)$ is a
four-component spinorial wave function and $\Gamma_{\mu}$ is the spin
connection. 
The only nonvanishing spin connection in this case is
$\Gamma_{\varphi}=i(1-\alpha) \sigma_{z}/2$, while the Dirac matrices are
conveniently defined as $\alpha^{i}=\gamma^{0}\gamma^{i}$,
$\beta =\gamma^{0}$ \cite{PRD.2008.78.64012,JoHEP.2004.2004.16}.

The magnetic flux tube in the background space described by the metric
above considered is related
\cite{JoHEP.2004.2004.16} to the magnetic field
$s \mathbf{B}=s(\nabla \times \mathbf{A}) =
\frac{s\overline{\phi}}{\alpha}\frac{\delta(r)}{r}\mathbf{\hat{z}}$
(where $\overline{\phi}=\phi /2\pi$ is the flux parameter),
while the vector potential in the Coulomb gauge
is $\mathbf{A}_{\varphi}=
\frac{\overline{\phi}}{\alpha r}\mathbf{\hat{\varphi}}$, 
with $s=\pm 1$ being twice the spin projection parameter. 
The parameter $s$ implies that the Dirac equation describes the planar
motion (in the absence of the $z$ coordinate) of the particle having only
one projection of three-dimensional spin vector.
To examine the physical implications  of these equations, we consider
their nonrelativistic limit. In this context, writing
$\psi=(\Phi,X)^{T}e^{-i M t}$ the Schr\"{o}dinger-Pauli equation is
$H\Phi=i \partial_{t}\Phi$, with
\begin{equation}
H=\frac{1}{2M}\left[ \frac{1}{i}\nabla_{\alpha }
\boldsymbol{-}\frac{q\overline{\phi}}{\alpha r}+
\frac{1-\alpha }{2\alpha r}\sigma
_{z}\right]^{2}-\frac{q s \overline{\phi}}{2M\alpha }\frac{\delta(r)}{r},
\end{equation}
where $\nabla_{\alpha }^{2}$ $=\frac{\partial^{2}}{\partial r^{2}}
+\frac{1}{r}\frac{\partial }{\partial r}+\frac{1}{\alpha^{2}r^{2}}
\frac{\partial^{2}}{\partial \varphi^{2}}$ is the Laplacian operator
in the conical space, and 
$\sigma_{i}=(\sigma_{r},\sigma_{\varphi},\sigma_{z})$
are the Pauli matrices in cylindrical coordinates.

For this system the total angular momentum operator,
$\hat{J}=-i\nabla_{\varphi }+\sigma_{z}/2$, commutes with the effective 
Hamiltonian. So, the solution to the Schr\"{ö}dinger-Pauli equation can be 
written in the form
\begin{equation}
\label{eq:wavef}
  \Phi(t,r,\varphi) =
  e^{-i\mathcal{E}t}
  \left(
  \begin{array}{c}
   f_{1}(r) e^{i(m-s/2) \varphi}\\
   f_{2}(r) e^{i(m+s/2) \varphi }
 \end{array}
 \right)
\end{equation}
with $m=n+1/2$, $n\in\mathbb{Z}$. At the same time, the radial equation
for $f_{1}(r)$ becomes
\begin{equation}
\label{eq:eigen}
\mathcal{H}f_1(r)=\mathcal{E}f_1(r),
\end{equation}
where
\begin{equation}
\label{eq:hfull}
\mathcal{H}=\mathcal{H}_{0}+\mathcal{U}_{\text{short}},
\end{equation}
\begin{equation}
\label{eq:hzero}
  \mathcal{H}_{0}=
  -\frac{1}{2M}
  \left[
    \frac{d^{2}}{dr^{2}}+
    \frac{1}{r}\frac{d}{dr}-\frac{j^{2}}{r^{2}}
  \right],
\end{equation}
\begin{equation}
\label{eq:ushort}
  \mathcal{U}_{\text{short}}=
  \frac{qs\overline{\phi}}{2M\alpha}\frac{\delta(r)}{r},
\end{equation}
with
$j=\frac{1}{\alpha}(m-\frac{s}{2}-q\overline{\phi}+\frac{1-\alpha}{2})$. 
The Hamiltonian in Eq. \eqref{eq:hfull} governs the quantum dynamics of
a spin-1/2 charged particle in the conical spacetime, with a magnetic 
field $\boldsymbol{B}$ along the z-axis, i.e., a spin-1/2 AB problem
in the conical space.
Let us consider a conical defect with a nucleus with radius
$r_{0}$, so it is suitable to write $\mathcal{U}_{\text{short}}(r)$ as
\cite{PRL.1990.64.503,PRD.1993.48.5935}
\begin{equation}
\label{eq:ushortr0}
\overline{\mathcal{U}}_{\text{short}}(r)=
\frac{qs\overline{\phi}}{2M\alpha} \frac{\delta (r-r_{0})}{r_{0}},
\end{equation}
and, at the end, the limit $r_{0}\to 0$ is taken. Although the
functional structure of  $\mathcal{U}_{\text{short}}$ and
$\overline{\mathcal{U}}_{\text{short}}$ are quite different, as
discussed in \cite{PRL.1990.64.503}, we are free to use any form
of potential provided that only the contribution of the form
\eqref{eq:ushort}  is excluded.

The operator $\mathcal{H}_{0}$, with domain
$\mathcal{D}(\mathcal{H}_{0})$, is self-adjoint if
$\mathcal{D}(\mathcal{H}_{0}^{\dagger})=\mathcal{D}(\mathcal{H}_{0})$
and $\mathcal{H}_{0}^{\dagger}=\mathcal{H}_{0}$.
For smooth functions, $g \in C_{0}^{\infty}(\mathbb{R}^2)$ with
$g(0)=0$, we should have $\mathcal{H}g=\mathcal{H}_{0}g$, and hence it
is reasonable to interpret the Hamiltonian \eqref{eq:hfull} as a
self-adjoint extension of
$\mathcal{H}_{0}|_{C_{0}^{\infty}(\mathbb{R}^{2}\setminus \{0\})}$ 
\cite{crll.1987.380.87,JMP.1998.39.47,LMP.1998.43.43}.
In order to proceed to the self-adjoint extensions of \eqref{eq:hzero},
we decompose the Hilbert space 
$\mathscr{H}=L^{2}(\mathbb{R}^{2})$
with respect to the angular momentum
$\mathscr{H}=\mathscr{H}_{r}\otimes\mathscr{H}_{\varphi}$, where
$\mathscr{H}_{r}=L^{2}(\mathbb{R}^{+},rdr)$ and
$\mathscr{H}_{\varphi}=L^{2}(\mathcal{S}^{1},d\varphi)$, with
$\mathcal{S}^{1}$ denoting the unit sphere in $\mathbb{R}^{2}$.
The operator $-\frac{\partial^{2}}{\partial\varphi^{2}}$ is essentially
self-adjoint in $L^{2}(\mathcal{S}^{1},d\varphi)$
\cite{Book.1975.Reed.II} and we obtain the operator $\mathcal{H}_{0}$ in
each angular momentum sector.
Now, using the unitary operator
$V:L^{2}(\mathbb{R}^{+},rdr)\to L^{2}(\mathbb{R}^{+},dr)$,
given by $(V g)(r)=r^{1/2}g(r)$, the operator
$\mathcal{H}_{0}$ becomes
\begin{equation}
  h_{0}=
  V\mathcal{H}_{0}V^{-1}=-\frac{1}{2M}
  \left[
    \frac{d^{2}}{dr^{2}}+\left(j^{2}-\frac{1}{4}\right)\frac{1}{r^{2}}
  \right],
\end{equation}
which is essentially self-adjoint for $|j| \geq 1$, while for $|j|< 1$ it
admits a one-parameter family of self-adjoint extensions
\cite{Book.1975.Reed.II}, $\mathcal{H}_{0,\lambda_{j}}$, where
$\lambda_{j}$ is the self-adjoint extension parameter.
To characterize this family, we will use the KS \cite{CMP.1991.139.103} 
and the BG \cite{JMP.1985.26.2520} approaches, both based in boundary
conditions.

In the KS approach, the boundary condition is a match of the logarithmic
derivatives of the zero-energy solutions for Eq. \eqref{eq:eigen}
and the solutions for the problem $\mathcal{H}_{0}$ plus self-adjoint
extension.
In the BG approach, the boundary condition is a mathematical limit
allowing divergent solutions of the Hamiltonian \eqref{eq:hzero} at
isolated points, provided they remain square integrable.

Now, the goal is to find the bound states for the Hamiltonian
\eqref{eq:hfull}. Following \cite{CMP.1991.139.103}, we temporarily
forget the $\delta $-function potential and find the boundary conditions
allowed for $\mathcal{H}_{0}$. 
But the self-adjoint extension provides infinity possible boundary
conditions, so that it cannot give us the true physics of the
problem. Nevertheless, once the physics at $r=0$ is known
\cite{AP.2010.325.2529,AP.2008.323.3150,PRD.1989.40.1346}, it is
possible to determine any arbitrary parameter coming from the
self-adjoint extension, so that it is possible to obtain a complete
description of the problem. 
Since we have a singular point, we must guarantee that the Hamiltonian is
self-adjoint in the region of motion. Note that  even if
$\mathcal{H}_{0}^{\dagger}=\mathcal{H}_{0}$, their domains could be
different.

We must find the deficiency subspaces, $\mathcal{N}_{\pm }$ , with
dimensions $n_{+}$ and $n_{-}$, respectively, which are called
deficiency indices of $\mathcal{H}_{0}$  \cite{Book.1975.Reed.II}. A
necessary and sufficient condition for $\mathcal{H}_{0}$ being
essentially self-adjoint is that $n_{+}=n_{-}=0$. On the other hand, if
$n_{+}=n_{-}\geq 1$, then $\mathcal{H}_{0}$ has an infinite number of
self-adjoint extensions parametrized by unitary operators
$U:\mathcal{N}_{+}\to\mathcal{N}_{-}$.

Next, we substitute the problem in Eq. \eqref{eq:eigen} by
$\mathcal{H}_{0}f_{\varrho}=\mathcal{E}f_{\varrho}$,
with $f_{\varrho }$ labeled by a parameter $\varrho$ which is related
to the behavior of the wave function in the limit $r\rightarrow
r_{0}$. But we cannot impose any boundary condition (e.g. $f=0$ at
$r=0$) without discovering which boundary conditions are allowed to
$\mathcal{H}_{0}$.
Then, from Eq. \eqref{eq:hzero} we achieve the modified Bessel equation
($\kappa^{2}=-2M\mathcal{E}$, $\mathcal{E}<0$)
\begin{equation}
\label{eq:extnew}
\left[
  \frac{d^{2}}{dr^{2}}+\frac{1}{r}\frac{d}{dr}-
  \left(\frac{j^{2}}{r^{2}}+\kappa^{2}\right) 
\right] f_{\varrho}(r) =0.
\end{equation}

Now, in order to find the full domain of $\mathcal{H}_{0}$ in
$L^{2}(\mathbb{R}^{+},rdr)$, we have to find its deficiency
subspace. To do this, we solve the eigenvalue equation
\begin{equation}
\label{eq:eigendefs}
  \mathcal{H}_{0}^{\dagger}f_{\varrho }^{\pm }
  =\pm i f_{\varrho }^{\pm },
\end{equation}
where $\mathcal{H}_{0}$ is given by Eq. \eqref{eq:hzero}.
The only square-integrable functions that are solutions of Eq.
\eqref{eq:eigendefs} are the modified Bessel functions
$K_{|j|}(r\sqrt{\mp \varepsilon })$, with $\varepsilon=2iM$. 
These functions are square integrable only in
the range $|j|<1$, for which $\mathcal{H}_{0}$ is not
self-adjoint. The dimension of such deficiency subspace is
$(n_{+},n_{-})=(1,1)$. Thus, 
$\mathcal{D}(\mathcal{H}_{0})$ in $L^{2}(\mathbb{R}^{+},rdr)$
is given by the set of functions \cite{Book.1975.Reed.II}
\begin{equation}
\label{eq:domain}
f_{\varrho }(r)=f_{1,j}(r)+
C\left[ K_{|j|}(r\sqrt{-\varepsilon})+
e^{i\varrho}K_{|j|}(r\sqrt{\varepsilon}) \right] ,
\end{equation}
where $f_{1,j}(r)$, with $f_{1,j}(r_{0})=\dot{f}_{1,j}(r_{0})=0$
($\dot{f}\equiv df/dr$), is the regular wave function when we do not have
$\overline{\mathcal{U}}_{\text{short}}(r)$.
The last term in Eq. \eqref{eq:domain}  gives the correct behavior for
the wave function when $r=r_{0}$. The parameter $\varrho ($mod$ 2\pi)$
represents a choice for the boundary condition. As we shall
see below, the physics of the problem determines such a parameter without 
ambiguity. In fact, $\varrho$ describes the coupling between
$\overline{\mathcal{U}}_{\text{short}}(r)$ and the wave function.
Thus, it must be expressed in terms of $\alpha $, the defect core radius  
$r_{0}$ and the effective angular momentum $j$. The next step is to find
a fitting for $\varrho$ compatible with 
$\overline{\mathcal{U}}_{\text{short}}(r)$.
In this sense, we write Eq. \eqref{eq:eigen} for  $\mathcal{E}=0$,
implying the zero-energy solution, $\mathcal{H}f_{0}=0$.
Now, we require the continuity for the logarithmic derivative 
\begin{equation}
  \label{eq:ftrue}
  \frac{\dot{f}_{0}}{f_{0}}
  \Big|_{r=r_{0}}=\frac{\dot{f}_{\varrho}}{f_{\varrho}}\Big|_{r=r_{0}},
\end{equation}
where $f_{\varrho}(r)$ comes from Eq. \eqref{eq:domain}.
However, since $r_{0}\approx 0$, the right-hand side of the Hamiltonian
\eqref{eq:ftrue} is calculated using the asymptotic representation for
Eq. \eqref{eq:domain} in the limit $r \to 0$. The left-hand side of
Eq. \eqref{eq:ftrue} is achieved integrating  the equation
$\mathcal{H}f_{0}=0$, from $0$ to $r_{0}$, which yields the parameter
$\varrho$ in terms of the physics of the problem, i.e., the correct
behavior of the wave functions for $r \to r_{0}$. 
By solving Eq. \eqref{eq:ftrue} for $\mathcal{E}$, we find the energy
spectrum 
\begin{equation}
  \label{eq:energy_KS}
  \mathcal{E}=
  -\frac{2}{Mr_{0}^{2}}
  \left[
    \frac{\Gamma (1+|j|)}{\Gamma (1-|j|)}
    \left(
      \frac{1 + \frac{\overline{\phi}}{\alpha |j|}+\frac{|j|}{2}}
           {1 - \frac{\overline{\phi}}{\alpha |j|}-\frac{|j|}{2}}
    \right)
   \right]^{1/|j|}.
\end{equation}
Notice that there is no arbitrary parameters in the above equation.

The above approach has the advantage of yielding the self-adjoint
extension parameter in terms of the physics of the problem, but it is 
not appropriate for dealing with scattering problems.
On the other hand, the BG method \cite{JMP.1985.26.2520} is suitable to 
address both bound and scattering scenarios, with the disadvantage of
allowing arbitrary self-adjoint extension parameters. Now, we apply the
BG approach to solve bound and scattering problems. By comparing the
results of these two approaches for bound states, the self-adjoint
extension parameter can be determined in terms of the physics of the
problem. Here, all self-adjoint extensions of
$\mathcal{H}_{0,\lambda_{j}}$ are parametrized by the boundary
condition at the origin \cite{JMP.1985.26.2520} 
($g_{0}(r)=\lim_{r\rightarrow 0^{+}}r^{|j|}\;g(r)$)
\begin{equation}
\label{eq:bc}
g_{0}(r)=\lambda_{j}\lim_{r\rightarrow0^{+}}
\frac{1}{r^{|j|}}\left[ g(r)-g_{0}(r')
\frac{1}{r^{|j|}}\right].
\end{equation}
The solutions for $\mathcal{H}_{0} f_{1,j}=k^{2} f_{1,j}$ 
($k^{2}=2M\mathcal{E}$) for $r \neq 0$, can be written as 
($\rho=2ikr$)
\begin{eqnarray}
\label{eq:general_sol}
f_{1,j}(r)
&=& \;A_{j }e^{-\frac{\rho}{2}} \rho^{|j|}
\;_{1}F_{1}\big(\frac{1}{2}+|j| ,1+2|j| ,\rho\big)\nonumber  \\
& &+ B_{j}e^{-\frac{\rho}{2}}\;\rho^{-|j|}
\;_{1}F_{1}\big( \frac{1}{2}-|j|
,1-2|j| ,\rho\big),\;\;\;\;\;\;\;\;
\end{eqnarray}
where $_{1}F_{1}(a,b,z)$ represents the confluent hypergeometric
function, and $A_{j}$, $B_{j}$ are the coefficients of the regular
and irregular solutions, respectively. By implementing 
Eq. \eqref{eq:general_sol} into the boundary condition \eqref{eq:bc}, we
derive the following relation between the coefficients $A_{j}$ and
$B_{j}$:
\begin{equation}
  \label{eq:coef_rel_1}
  \lambda_{j}A_{j}=(2ik)^{-2|j|}B_{j}
  \bigg(
  1+\frac{\lambda_{j}k^{2}}{4(1-|j|)}
  \lim_{r\rightarrow 0^{+}}r^{2-2|j|}
  \bigg).
\end{equation}
In the above equation, the coefficient of $B_{j }$ diverges as
$\lim_{r\rightarrow 0^{+}}r^{2-2|j|}$, if $|j|>1.$
Thus, $B_{j}$ must be zero for $|j|>1$, and the
condition for the occurrence of a singular solution is $|j|< 1$. So, the 
presence of an irregular solution stems from the fact the
operator is not self-adjoint for $|j|<1$, and this irregular solution is
associated with a self-adjoint extension of the operator
$\mathcal{H}_{0}$  \cite{JPA.1995.28.2359,PRA.1992.46.6052}.
In other words, the self-adjoint extension essentially consists in
including irregular solutions in $\mathcal{D}(\mathcal{H}_{0})$, which
allows us to select an appropriate boundary condition for the problem.

In the present system the energy of a bound state has to be negative,
so that $k$ is a pure imaginary, $k = i \kappa$. Thus, with the
substitution $k \to i \kappa$, we have ($\rho'=-2\kappa r$) 
\begin{eqnarray}
\label{eq:bound}
f_{1,j}^{\mathcal{B}}(r)&=& A_{j} \;e^{-\frac{\rho'}{2}}\; {\rho'}^{|j|}\;
_{1}F_{1}\big(\frac{1}{2}+|j|,1+2|j|,\rho'\big)\nonumber\\
& &+ B_{j}\; e^{-\frac{\rho'}{2}}\, {\rho'}^{-|j|} \;
_{1}F_{1}\big(\frac{1}{2}-|j|,1-2|j|,\rho'\big).\;\;\;\;\;\;\;\;
\end{eqnarray}
For Eq. \eqref{eq:bound} representing a bound state, the solution
$f_{1,j}^{\mathcal{B}}(r)$ must vanish for $r\to \infty$, i.e.,
it must be normalizable. By using the asymptotic representation of
$_{1}F_{1}(a,b,z)$ for $r\to \infty$, the normalizability condition
yields the relation
\begin{equation} 
  \label{eq:coef_rel_2}
  B_{j}=-16^{|j|} \frac{\Gamma(1+|j|)}{\Gamma(1-|j|)} A_{j}.
\end{equation}
From Eq. \eqref{eq:coef_rel_1}, for $|j| < 1$ we have
$B_{j} =\lambda_{j} (-2\kappa)^{2 |j|} A_{j}$;
and by using Eq. \eqref{eq:coef_rel_2}, the bound state energy is
\begin{equation}
  \label{eq:energy_BG}
  \mathcal{E}=-\frac{2}{M} 
  \left[
    -\frac{1}{\lambda_{j}}
    \frac{\Gamma(1+|j|)}{\Gamma(1-|j|)} 
  \right]^{1/|j|}.
\end{equation}
This coincides with Eq. (3.13) of Ref. \cite{PRD.1994.50.7715}
for $\alpha=1$,
i.e., the spin-1/2 AB problem in Euclidean space
with the spinorial connection. 
By comparing
Eq. \eqref{eq:energy_BG} with Eq. \eqref{eq:energy_KS}, we find
\begin{equation}
  \label{eq:lambdaj}
  \frac{1}{\lambda_{j}}=
  -\frac{1}{r_{0}^{2 |j|}}
  \left(
    \frac{1 + \frac{\overline{\phi}}{\alpha |j|}+\frac{|j|}{2}}
         {1 - \frac{\overline{\phi}}{\alpha |j|}-\frac{|j|}{2}}
  \right).
\end{equation}
We have thus attained a relation between the self-adjoint extension
parameter and the physical parameters of the problem, $j$ and
$r_{0}$. It should be mentioned that some relations involving the
self-adjoint extension parameter and the $\delta$-function coupling
constant were previously obtained by using Green's function in Ref.
\cite{JMP.1995.36.5453} and the renormalization technique in Ref.
\cite{Book.1995.Jackiw}, being both, however, deprived from a clear
physical interpretation.

Once the bound energy problem has been examined, let us now analyze the
AB scattering scenario. In this case, the boundary condition is again
given by Eq. \eqref{eq:bc} but now with the replacement
$\lambda_{j}\to\lambda_{j}^{s}$, where $\lambda_{j}^{s}$ is the
self-adjoint extension parameter for the scattering problem. In the
scattering analysis it is more convenient to use the solution of the
equation $\mathcal{H}_{0} f_{1,j}=k^{2} f_{1,j}$, in terms of Bessel
functions 
\begin{equation}
\label{eq:sol1}
f_{1,j}(r) = C_{j} J(|j|,k r) + D_{j} Y(|j|, k r),
\end{equation}
with $C_{j}$ and $D_{j}$ being constants. Upon replacing $f_{1,j}(r)$ in 
the boundary condition \eqref{eq:bc}, we obtain
$\lambda_{j}^{s} C_{j} \xi k^{|j|} = D_{j} \big[\zeta k^{-|j|}
-\lambda_{j}^{s} (\eta k^{|j|}+\zeta\gamma
k^{-|j|}\lim_{r\to 0^{+}} r^{2-2|j|})\big]$,
where $\xi = \frac{1}{2^{|j|}\Gamma(1+|j|)}$, 
$\zeta = -\frac{2^{|j|}\Gamma(|j|)}{\pi}$, 
$\eta = - \frac{\cos(\pi |j|)\Gamma (-|j|)}{\pi 2^{|j|}}$ and 
$\gamma = \frac{k^2}{4(1-|j|)}$.
As in the bound state calculation, whenever $|j|<1$, we have $D_{j}\neq
0$; this means that there arises again the contribution of the irregular 
solution $Y$ at the origin when the operator is not self-adjoint. Thus, for
$|j|<1$, we obtain
$\lambda_{j}^{s} C_{j} \xi k^{|j|}=
D_{j}(\zeta k^{-|j|}-\lambda_{j}^{s}\eta k^{|j|})$,
and by substituting the values of $\xi$, $\zeta$ and $\eta$ into above
expression we find $D_{j} = -\mu_{j}^{\lambda_{j}^{s}} C_{j}$, where
\begin{equation}  
  \label{eq:mul}
  \mu_{j}^{\lambda_{j}^{s}} = \frac{\lambda_{j}^{s} k^{2|j|}
    \Gamma(1-|j|)\sin(|j|\pi)} {\lambda_{j}^{s} k^{2|j|} \Gamma{(1-|j|)}
    \cos(\pi |j|) + 4^{|j|} \Gamma(1+|j|)}.
\end{equation}
Since the $\delta$ is a short range potential, it follows that the
behavior of $f_{1,j}$ for  $r\to\infty$ is given by
\begin{equation}
  \label{eq:f1asim}
  f_{1,j}(r)\sim \sqrt{\frac{2}{\pi k r}}
  \cos\big[k r - \frac{1}{2}|m|\pi-\frac{\pi}{4}+
  \delta_{j}^{{\lambda_{j}^{s}}}(k,\overline{\phi})\big],
\end{equation}
where $\delta_{j}^{{\lambda_{j}^{s}}}(k,\overline{\phi})$ 
is a scattering phase shift.
The phase shift is a measure of the argument difference to the
asymptotic behavior of the solution $J(|m|,k r)$  of the radial-
free equation  that is regular at the origin. By using the asymptotic
behavior of $J(|j|,k r)$ and $Y(|j|,k r)$ for $r\to\infty$ in Eq.
\eqref{eq:sol1}, and comparing it with Eq. \eqref{eq:f1asim}, similarly
as  done in \cite{JPA.2010.43.354011}, we found that
$\delta_{j}^{{\lambda_{j}^{s}}}(k,\overline{\phi}) 
= \Delta_m(\overline{\phi}) +\theta_{{\lambda_{j}^{s}}}$,
where $\Delta_m(\overline{\phi}) =\frac{\pi}{2}(|m|-|m+\overline{\phi}|)$,
and $\theta_{{\lambda_{j}^{s}}}=\arctan{(\mu_{j}^{\lambda_{j}^{s}})}$.
Therefore, the expression for the $S$ matrix is
\begin{equation}
  \label{eq:smatrix}
  S= e^{2 i \Delta_m(\overline{\phi})}
  \bigg[
  \frac{\lambda_{j}^{s} k^{2 |j|}\Gamma(1-|j|)e^{i|j|\pi}+4^{|j|}\Gamma{(1+|j|)}}
  {\lambda_{j}^{s} k^{2|j|}\Gamma(1-|j|)e^{-i|j|\pi}+4^{|j|}\Gamma{(1+|j|)}}
  \bigg].
\end{equation}
In accordance with the general theory of scattering, the poles of the
$S$ matrix in the upper half of the complex plane
\cite{PRC.1999.60.34308} [these poles occur in the denominator of
\eqref{eq:smatrix} with the replacement $k \to i \kappa $] determines
the positions of the bound states in the energy scale,
Eq. \eqref{eq:energy_BG}.
From this, we have $\lambda_{j}^{s}=\lambda_{j}$, with $\lambda_{j}$
given by Eq. \eqref{eq:lambdaj}, and the self-adjoint extension parameter
for the scattering scenario being the same as that for the bound
state problem. 
This is a very interesting result that has not been described
in the literature yet, as far as we know. Thus, we also obtain the phase 
shift and the scattering matrix in terms of the physics of the problem.
If ${\lambda_{j}^{s}} = 0$, we achieve the corresponding result for the
pure AB problem with the Dirichlet boundary condition; in this case, we
recover the expression for the scattering matrix found in
Ref. \cite{AP.1983.146.1}, $S= e^{2 i\Delta_m(\overline{\phi})}$. If we make
${\lambda_{j}^{s}} = \infty$, we get $S=e^{2i\Delta_m(\overline{\phi})+ 2i\pi |j|}$.

In this article, we have presented a general regularization method to 
address a system endowed with a singular Hamiltonian (due to localized
fields sources or quantum confinement). Using the KS approach, the bound
states were determined in terms of the physics of the problem, in a
very consistent way and without any arbitrary parameter.
In sequel, we employed the BG approach; by comparing the results of these
approaches, we have determined the value of the self-adjoint extension
parameter for the bound state problem, which coincides with the one for
scattering problem. We thus obtain the $S$ matrix in terms of the physics
of the problem, as well. A natural extension of the problem studied
here, amongst many possible options, is the inclusion of the Coulomb
potential, which naturally appears in two-dimensional systems, such
as graphene \cite{RMP.2009.81.109} and anyons  systems
\cite{PRL.1982.49.957,IJMP.1989.B3.1001}. Results in this respect will
be reported elsewhere. 

The authors would like to thank M. G. E. da Luz, D. Bazeia, E. R. Bezerra 
de Mello, C. R. de Oliveira and M. M. Ferreira Jr. for their critical
reading of the manuscript, and for helpful discussions and encouragement. 
E. O. Silva acknowledges research grants from FAPEMA, CAPES-(PNPD)
and M. Pereira acknowledges research grants from Funda\c{c}\~{a}o 
Arauc\'{a}ria.

\end{document}